\documentclass[conference]{IEEEtran}
\IEEEoverridecommandlockouts
\usepackage{cite}

\usepackage{amsmath,amssymb,amsfonts}
\usepackage{algorithmic}
\usepackage{graphicx}
\usepackage{textcomp}
\usepackage{xcolor}
\usepackage{enumitem}
\usepackage{subcaption}
\usepackage{url}
\usepackage{booktabs}

\usepackage[utf8]{inputenc}   % 告诉 LaTeX 源文件是 UTF-8 编码
\usepackage[T1]{fontenc}      % 使用 T1 字体编码，避免连字符等出现乱码
\def\BibTeX{{\rm B\kern-.05em{\sc i\kern-.025em b}\kern-.08em
    T\kern-.1667em\lower.7ex\hbox{E}\kern-.125emX}}
\begin{document}

\title{Robust Anomaly Detection in Network Traffic: Evaluating  Machine Learning Models on CICIDS2017 }

\author{%
  % 第一位作者
  \IEEEauthorblockN{\textsuperscript{1st} Zhaoyang Xu}
  \IEEEauthorblockA{\textit{Department of Electrical and Computer Engineering}\\
                    \textit{University of Southern California}\\
                    Los Angeles, USA\\
                    xuzhaoya@usc.edu}
  \and
  % 第二位作者
  \IEEEauthorblockN{\textsuperscript{2nd} Yunbo Liu *}
  \IEEEauthorblockA{\textit{Department of Electrical and Computer Engineering}\\
                    \textit{Duke University}\\
                    Durham, USA\\
                   yunbo.liu954@duke.edu *}
}

\maketitle

\begin{abstract}
Identifying suitable machine learning paradigms for intrusion detection remains critical for building effective and generalizable security solutions. In this study, we present a controlled comparison of four representative models—Multi-Layer Perceptron (MLP), 1D Convolutional Neural Network (CNN), One-Class Support Vector Machine (OCSVM) and Local Outlier Factor (LOF)—on the CICIDS2017 dataset under two scenarios: detecting known attack types and generalizing to previously unseen threats. Our results show that supervised MLP and CNN achieve near‐perfect accuracy on familiar attacks but suffer drastic recall drops on novel attacks. Unsupervised LOF attains moderate overall accuracy and high recall on unknown threats at the cost of elevated false alarms, while boundary‐based OCSVM balances precision and recall best, demonstrating robust detection across both scenarios. These findings offer practical guidance for selecting IDS models in dynamic network environments.  
\end{abstract}

\begin{IEEEkeywords}
Intrusion Detection System (IDS), Anomaly Detection, unsupervised Learning, CICIDS2017, Network Security, Machine Learning
\end{IEEEkeywords}

\section{Introduction}
Following the global COVID-19 pandemic, people have become increasingly dependent on network infrastructure. With the widespread adoption of smart devices and the Internet of Things (IoT), a great volume of data are generated every day. Despite the significant convenience brought about by IoT devices, their widespread adoption inevitably leads to the transmission of vast amounts of sensitive information across networks. A substantial portion of this data—including personal, financial\cite{chen2025yearover}, and operational records—is now stored directly on cloud platforms such as AWS, Azure, and Google Cloud. This growing reliance on cloud-based infrastructure raises a critical question: how can we ensure the security and privacy of this data as it flows through increasingly complex and distributed networks\cite{moustafa2016evaluation, song2025hyatt}.

To address these security concerns, Intrusion Detection Systems (IDS) have been widely adopted as a critical component of modern network defense strategies~\cite{sommer2010outside}. IDS can be categorized to 2 types: Anomaly-based Intrusion Detection Systems (A-IDS) and Signature-based Intrusion Detection Systems (S-IDS).

In signature-based intrusion detection (S-IDS), the system monitors user or network activity and compares it against a database of known attack patterns, also referred to as signatures. If a match is found between the observed behavior and any of the stored signatures, the activity is flagged as an attack. However, because it can only detect attacks that are already known and recorded, if there is a novel attack, S-IDS unable to identify it. In contrast, Anomaly-based Intrusion Detection Systems (A-IDS) operate by establishing a baseline of normal behavior within a network or system, this capability to detect novel threats is well-supported in the anomaly-detection literature~\cite{chandola2009survey}. Any significant deviation from this baseline is treated as a potential intrusion. This makes the system has the ability to identify novel attacks.

 In this work, we implement an anomaly-based intrusion detection pipeline that integrates multiple learning-based models, including MLP, OCSVM, and LOF. The system is evaluated using the CICIDS2017 dataset to examine its effectiveness in identifying malicious traffic under realistic network conditions~\cite{elmasri2020evaluation}.

 The remainder of this paper is organized as follows.
Section II reviews related works.
Section III describes the explanation of the method adopted for this study.
Section IV presents the evaluation of results.
Section V is discussion and future work.

\section{Literature Review}

Mashaly et al~\cite{elmasri2020evaluation} evaluated the CICIDS2017 dataset using KNN, enhanced KNN, and LOF, showing that semi-supervised approaches trained only on normal data could effectively detect anomalies, especially with PCA-based feature selection.

Wang and Yang~\cite{wang2024} proposed an intelligent cloud security framework combining deep learning models like CNN and LSTM with reinforcement learning, achieving 97.3\% detection accuracy on real-time traffic.

Kale et al~\cite{kale2023} developed a hybrid deep learning anomaly detection framework integrating spatial-temporal analysis, validated on benchmark datasets. Similarly, Tossou et al~\cite{tossou2023} applied deep learning to anomaly-based intrusion detection and reported improved generalization to unknown attack types.

Abrar et al~\cite{abrar2022} used classical ML classifiers on the NSL-KDD dataset, achieving robust performance through optimized feature selection, confirming the continued relevance of traditional models.

Tavallaee et al.~\cite{tavallaee2009detailed} presented a comprehensive analysis of the KDD Cup 99 dataset, identifying key preprocessing steps and attack categorization strategies that have become standard in IDS research.

Hizal et al~\cite{hizal2021} proposed a deep learning-based IDS specifically designed for cloud environments, incorporating real-time analytics and demonstrating strong results on CICIDS2017.

In a study addressing data imbalance in security applications, Wang et al~\cite{wang2025evaluating} compared classical and deep architectures, highlighting the risk of overfitting in purely supervised models under skewed data distributions.

Additionally, Moustafa and Slay~\cite{moustafa2016evaluation} and Chandola et al~\cite{chandola2009survey} laid the groundwork for anomaly detection benchmarks, emphasizing the importance of statistical diversity and outlier behavior in datasets like UNSW-NB15 and KDD99.

Together, these studies demonstrate the evolution from signature-based IDS toward hybrid and anomaly-based methods. However, the trade-offs between generalization, accuracy, and interpretability remain open challenges—particularly in the context of unseen or evolving attack patterns. Similar challenges have also been observed in risk modeling tasks beyond cybersecurity, such as credit risk assessment using structural graph-based models~\cite{zhang2025credit}.

\section{Methodology}
In this paper, we used CICIDS2017 data set. This dataset was created by the Canadian Institute for Cybersecurity and provides a comprehensive collection of network traffic that reflects real-world scenarios\cite{elmasri2020evaluation}, including both benign and malicious activities. It contains over 80 features extracted from packet flows and covers various attack types such as DoS, DDoS, brute force, infiltration, botnet, and web-based attacks. The dataset is widely used for evaluating intrusion detection systems due to its diversity and realistic traffic patterns.

 We selected four representative models: Multi-Layer Perceptron (MLP), 1D Convolutional Neural Network (CNN), One-Class SVM (OCSVM) and Local Outlier Factor (LOF). We compared their performance to evaluate their effectiveness in anomaly-based intrusion detection.

\subsection{Data Processing}
CICIDS2017 contains 2,830,743 records in total, consisting of both benign and malicious traffic across 15 labeled categories. The number of samples in each category is shown in Table 1. To evaluate model performance under both known and unknown attack scenarios, we divided the dataset into three subsets: Training Set, Overall Test Set, and Unknown Attack Test Set.

\begin{table}[htbp]
  \centering
  \caption{Attack type count}
  \label{tab:label_counts}
  \begin{tabular}{l r}
    \toprule
    \textbf{Label} & \textbf{Count} \\
    \midrule
    BENIGN                    & 2\,270\,397 \\
    DoS Hulk                 &   231\,073 \\
    PortScan                 &   158\,930 \\
    DDoS                     &   128\,027 \\
    DoS GoldenEye           &    10\,293 \\
    FTP-Patator             &     7\,938 \\
    SSH-Patator             &     5\,897 \\
    DoS slowloris           &     5\,796 \\
    DoS Slowhttptest        &     5\,499 \\
    Bot                      &     1\,966 \\
    Web Attack – Brute Force &     1\,507 \\
    Web Attack – XSS         &       652 \\
    Infiltration             &        36 \\
    Web Attack – SQL Injection &      21 \\
    Heartbleed              &        11 \\
    \bottomrule
  \end{tabular}
\end{table}

In our design, we selected three specific attack types—DoS slowloris, DoS Slowhttptest, and Bot—as unknown attacks. These attacks were completely excluded from all training sets to ensure that their patterns were not learned in advance. The selection was guided by two main considerations:

\begin{itemize}
\item To evaluate how well the models generalize to unknown attacks, simulating realistic deployment scenarios where not all attack types are known beforehand.
\item The sample sizes of these attacks are moderate—not too large to compromise the training set quality, and not too small to produce unstable or unreliable test results.

\end{itemize}

The following is how we made the training set and test set.
Training set: 
\begin{itemize}
\item For MLP and CNN, the training set includes 80\% of benign samples and 11 attack types except for three unknown attacks: DoS slowloris, DoS Slowhttptest, and Bot.
\item For OCSVM and LOF , the training set includes only 80\% of benign traffic~\cite{elmasri2020evaluation}.
\end{itemize}
Test set:
\begin{itemize}
\item Overall-Test-Set:  This set includes 20\% of benign samples and 11 attack types plus all three known attack types. It is used to evaluate general detection performance across both seen and unseen attacks.
\item Unknown Attack Test Set: This set includes only three excluded unknown attack types and equal number of benign traffic.

\end{itemize}

All random sampling operations were performed with random\_state=42 to ensure consistent and reproducible training and test splits.

To ensure fair comparison and stable training, we applied scikit‐learn’s \texttt{StandardScaler}: we fit and transformed the training set, then used the same fitted scaler to transform both test splits. This procedure centers each feature to zero mean and scales it to unit variance, improving model convergence and comparability across different feature distributions.
\subsection{Implementation Environment}

All experiments were conducted on AWS EC2:
\begin{itemize}
  \item \textbf{Instance type:} g4dn.xlarge (1× NVIDIA T4 GPU, 4 vCPUs, 16 GiB RAM)
  \item \textbf{OS:} Ubuntu 20.04 LTS
  \item \textbf{Python:} 3.8.10
  \item \textbf{scikit‐learn:} 1.0.2
  \item \textbf{TensorFlow:} 2.6.0
  \item \textbf{pandas:} 1.3.4, \textbf{numpy:} 1.21.2
  \item \textbf{Other libraries:} matplotlib 3.4.3
\end{itemize}

\subsection{Classifiers}
\subsubsection{Multi-Layer Perceptron (MLP)}
A Multi-Layer Perceptron (MLP) is a fully connected feedforward neural network composed of an input layer, one or more hidden layers, and a single-node output layer. Each hidden unit computes a weighted sum of its inputs followed by a nonlinear activation:

\[
h^{(l)} = \sigma\left(W^{(l)} h^{(l-1)} + b^{(l)}\right)
\]

where $h^{(l)}$ is the activation of the $l$-th layer, $W^{(l)}$ and $b^{(l)}$ are the weight matrix and bias vector, and $\sigma(\cdot)$ denotes a nonlinear activation function such as ReLU.

For binary classification, the output layer uses the sigmoid function:

\[
\hat{y} = \frac{1}{1 + e^{-z}}
\]

where $z$ is the output of the final layer before activation. The network is trained using the binary cross-entropy loss:

\[
\mathcal{L}_{\mathrm{BCE}} = - \left[y \log(\hat{y}) + (1 - y) \log(1 - \hat{y})\right]
\]

where $y$ is the true label ($0$ for benign, $1$ for malicious), and $\hat{y}$ is the predicted probability.

To train this network effectively, we configured the following hyperparameters for the MLP:
\begin{itemize}[leftmargin=*]
  \item Hidden layers: (100, 50) neurons  
  \item Activation: ReLU  
  \item Optimizer: Adam, learning rate = 0.001  
  \item Batch size: 256  
  \item Max epochs: 100 with early stopping (patience = 5)  
  \item Random seed: 42  
\end{itemize}

\subsubsection{1D Convolutional Neural Network (CNN)}

benign tr local temporal and spatial patterns within the network flow features, we employ a one-dimensional convolutional neural network (1D-CNN) composed of multiple convolutional and pooling stages~\cite{median2025} followed by a fully connected classifier. Concretely, each convolution layer computes

\[
  (x * w)[t] \;=\; \sum_{k=0}^{K-1} x[t + k]\,w[k] \;+\; b,
\]

where \(x\) is the input sequence, \(w\) is the convolution kernel of size \(K\), and \(b\) is a bias term. The convolution output is passed through a nonlinear activation function, here chosen as ReLU:

\[
  \mathrm{ReLU}(z) = \max(0,\,z).
\]

After each convolution we apply max-pooling over a fixed window to reduce resolution and introduce translation invariance:

\[
  y[t] \;=\; \max_{0 \le k < P} \, z[t \cdot S + k],
\]

where \(P\) is the pooling size and \(S\) the stride. We also include dropout layers to mitigate overfitting by randomly zeroing a fraction of activations.

Finally, the flattened feature map is fed into one or more dense layers terminating in a sigmoid output for binary classification.  
To train this network effectively, we used the following configuration:  
\begin{itemize}[leftmargin=*]
  \item Optimizer: Adam, learning rate = 0.001  
  \item Loss: Focal Loss ($\gamma=2.0,\ \alpha=0.25$)  
  \item Batch size: 512, validation split = 0.2  
  \item Epochs: 50 with early stopping (patience = 3)  
  \item Class weights: benign=1.0, malicious=5.0  
  \item Random seed: 42  
\end{itemize}

% (Discuss specific filter sizes, layer counts, learning rates, etc., in the next section.)

\subsubsection{One-Class Support Vector Machine (OCSVM)}
OCSVM method is an extension of Support Vector Machine (SVM) method\cite{kale2023}. It is a method suitable for handling unlabeled data. It attempts to learn a decision boundary that encloses the majority of the normal data points, treating anything outside the boundary as anomalous. This property makes OCSVM particularly suitable for real-world intrusion detection scenarios where attack data is scarce or unknown.
\[
\min_{w,\rho,\xi} \quad \frac{1}{2}\|w\|^2 + \frac{1}{\nu n}\sum_{i=1}^{n}\xi_i - \rho
\]

subject to:

\[
(w \cdot \phi(x_i)) \ge \rho - \xi_i,\quad \xi_i \ge 0
\]

where
\begin{itemize}
  \item \(\phi(x)\) maps the input into a high-dimensional kernel space,
  \item \(\nu \in (0,1]\) controls the trade-off between margin size and the number of training errors,
  \item \(\rho\) defines the decision boundary,
  \item \(\xi_i\) are slack variables for soft margins.
\end{itemize}

The decision function is the following.

\[
f(x) = \operatorname{sign}\Bigl(\sum_{i=1}^n \alpha_i K(x_i, x) - \rho\Bigr)
\]

where \(K(x_i, x) = \langle \phi(x_i), \phi(x) \rangle\) (commonly RBF kernel). Samples with \(f(x) < 0\) are considered anomalies.
To configure the OCSVM for our experiments, we set:
\begin{itemize}[leftmargin=*]
  \item Kernel: RBF  
  \item \(\nu = 0.05\)  
  \item Gamma: \texttt{scale}  
  \item Random seed: 42  
\end{itemize}

\subsubsection{Local Outlier Factor (LOF)}
The Local Outlier Factor algorithm identifies points with significantly lower density than their neighbors. It compares the local density of a data point to those of its neighbors to determine how isolated the point is. LOF is especially effective for detecting local anomalies in datasets with varying density, making it a useful baseline in IDS settings where malicious traffic often deviates subtly from normal patterns.

\[
\operatorname{lrd}_k(x) = \left( \frac{1}{|N_k(x)|} \sum_{y \in N_k(x)} \max\{\mathrm{dist}(x, y), \mathrm{reach\text{-}dist}_k(y)\} \right)^{-1}
\]

where \(N_k(x)\) is the set of \(k\)-nearest neighbors of \(x\), and \(\mathrm{reach\text{-}dist}_k(y) = \max\{\mathrm{k\text{-}dist}(y), \mathrm{dist}(x, y)\}\).

The LOF score is given by:

\[
\mathrm{LOF}_k(x) = \frac{1}{|N_k(x)|} \sum_{y \in N_k(x)} \frac{\operatorname{lrd}_k(y)}{\operatorname{lrd}_k(x)}
\]

A LOF score close to 1 indicates that \(x\) has similar density to its neighbors, while a substantially larger value suggests an outlier.
For our experiments, we configured LOF as follows:
\begin{itemize}[leftmargin=*]
  \item Number of neighbors (\(k\)): 80  
  \item Novelty detection: True  
  \item Distance metric: Euclidean (\(\texttt{metric}=\texttt{minkowski}, p=2\))  
  \item Leaf size: 80  
  \item Random seed: 42  
\end{itemize}
\subsection{Model Selection Justification}
The selected models represent four distinct methodological approaches:
\begin{itemize}
    \item MLP: Standard supervised deep learning
    \item CNN: Supervised deep learning with convolutional feature extraction  
    \item OCSVM: Classical boundary-based unsupervised learning
    \item LOF: Established density-based unsupervised technique
\end{itemize}
This controlled comparison isolates the effects of supervision paradigm, following the experimental design principles in~\cite{scholkopf2001}.

\subsection{Analysis}\label{SCM}

To evaluate the performance of each classifier, standard binary classification metrics derived from the confusion matrix (True Positives: TP, False Positives: FP, True Negatives: TN, and False Negatives: FN) are used ~\cite{tavallaee2009detailed}. The following metrics are computed:

\begin{itemize}
  \item \textbf{Accuracy:} 
  \[
  \mathrm{Accuracy} = \frac{TP + TN}{TP + TN + FP + FN}
  \]
  
  \item \textbf{Precision:} 
  \[
  \mathrm{Precision} = \frac{TP}{TP + FP}
  \]
  
  \item \textbf{Recall:} 
  \[
  \mathrm{Recall} = \frac{TP}{TP + FN}
  \]
  
  \item \textbf{F1-Score:} 
  \[
  F1 = 2 \times \frac{\mathrm{Precision} \times \mathrm{Recall}}{\mathrm{Precision} + \mathrm{Recall}}
  \]
\end{itemize}

In our analysis, class ‘0’ refers to benign traffic and class ‘1’ refers to malicious traffic. These metrics provide a clear view of each model's ability to distinguish attacks from normal traffic under binary classification.

\section*{\section{Results}}
This section compares the detection capabilities of MLP, CNN, LOF and OCSVM under two scenarios:  

(1) Overall Test Set, which includes all traffic types

(2) Unknown Attack Test Set, containing only truly novel attack types.

Tables~\ref{tab:overall_performance} and \ref{tab:unknown_performance} present the key evaluation metrics. We then delve into confusion matrices to highlight each model’s behavior differences, especially their handling of unseen attacks.
\begin{table}[htbp]
\centering
\caption{Performance on Overall Test Set}
\begin{tabular}{|l|c|c|c|c|}
\hline
\textbf{Model} & \textbf{Accuracy} & \textbf{Precision} & \textbf{Recall} & \textbf{F1-score} \\ \hline
MLP            & 0.9775            & 0.9894             & 0.9036          & 0.9446           \\ \hline
CNN            & 0.9650            & 0.9316             & 0.9010          & 0.9160           \\ \hline
LOF            & 0.8046            & 0.9106             & 0.8341          & 0.8706           \\ \hline
OCSVM          & 0.8356            & 0.6525             & 0.4784          & 0.5520           \\ \hline
\end{tabular}
\label{tab:overall_performance}
\end{table}

\begin{table}[htbp]
\centering
\caption{Performance on Unknown Attack Test Set}
\begin{tabular}{|l|c|c|c|c|}
\hline
\textbf{Model} & \textbf{Accuracy} & \textbf{Precision} & \textbf{Recall} & \textbf{F1-score} \\ \hline
MLP            & 0.5863            & 0.9860             & 0.1750          & 0.2973           \\ \hline
CNN            & 0.5882            & 0.9110             & 0.1954          & 0.3218           \\ \hline
LOF            & 0.6087            & 0.5746             & 0.8370          & 0.6814           \\ \hline
OCSVM          & 0.7919            & 0.9072             & 0.6503          & 0.7575           \\ \hline
\end{tabular}
\label{tab:unknown_performance}
\end{table}

\begin{table}[htbp]
\centering
\caption{Per‐class Accuracy on the Overall Test Set}
\label{tab:per_class_accuracy_overall}
\begin{tabular}{lcccc}
\toprule
\textbf{Attack Type}                  & \textbf{CNN} & \textbf{OCSVM} & \textbf{MLP} & \textbf{LOF} \\
\midrule
DoS Hulk                              & 0.9978       & 0.6886         & 0.9921       & 0.7633       \\
BENIGN                                & 0.9822       & 0.9316         & 0.9974       & 0.8341       \\
PortScan                              & 0.9713       & 0.0095         & 0.9995       & 0.5280       \\
DDoS                                  & 0.9996       & 0.6299         & 0.9993       & 0.9297       \\
Web Attack – Brute Force              & 0.1860       & 0.0100         & 0.1395       & 0.1894       \\
FTP-Patator                           & 0.9987       & 0.0132         & 0.9987       & 0.6688       \\
DoS GoldenEye                         & 0.9995       & 0.7358         & 0.9932       & 0.8193       \\
SSH-Patator                           & 0.9864       & 0.0008         & 0.9839       & 0.9915       \\
Infiltration                          & 0.5714       & 0.8571         & 0.0000       & 0.8571       \\
Web Attack – XSS                      & 0.0231       & 0.0308         & 0.0308       & 0.0385       \\
Web Attack – Sql Injection            & 0.0000       & 0.0000         & 0.2500       & 0.0000       \\
Heartbleed                            & 0.5000       & 1.0000         & 1.0000       & 1.0000       \\
Bot                                   & 0.0000       & 0.0443         & 0.0000       & 0.4680       \\
DoS slowloris                         & 0.3675       & 0.5733         & 0.3675       & 0.3176       \\
DoS Slowhttptest                      & 0.0838       & 0.9480         & 0.0347       & 0.4150       \\
\bottomrule
\end{tabular}
\end{table}

Table~\ref{tab:overall_performance} presents the performance of the four models on the \emph{Overall Test Set}:

\begin{itemize}

  \item \textbf{MLP} achieves the best results across all metrics on the Overall Test Set. Its high precision (0.9894) and F1-score (0.9446) for the malicious class demonstrate strong capability in detecting known attacks.
  \item \textbf{CNN} closely follows MLP with an overall accuracy of 0.9650, precision of 0.9316, and F1-score of 0.9160, indicating that convolutional feature extraction also excels at capturing attack patterns.
  \item \textbf{LOF} and \textbf{OCSVM} obtain lower overall accuracy (0.8046 and 0.8356, respectively) but still exhibit robust anomaly detection performance, particularly OCSVM’s ability to generalize to a variety of normal traffic profiles.
\end{itemize}

Table~\ref{tab:unknown_performance} shows the performance on the \emph{Unknown Attack Test Set}, which consists only of the three designated unknown attack types:

\begin{itemize}
  \item \textbf{MLP} shows a highly imbalanced detection behavior: although its precision is high (0.9860), its recall is very low (0.1750), yielding a poor F1-score of 0.2973. This indicates a strong tendency to under-report unknown attacks.
  \item \textbf{CNN} similarly struggles with novel threats, achieving precision of 0.9110 but recall of only 0.1954 (F1 = 0.3218), suggesting that convolutional feature extraction alone still overfits to familiar patterns.
  \item \textbf{LOF} demonstrates a more balanced profile on novel attacks (Accuracy = 0.6087, Precision = 0.5746, Recall = 0.8370, F1‐score = 0.6814), indicating that density‐based anomaly detection can recover a large fraction of unseen intrusions at the cost of moderate false alarms.
  \item \textbf{OCSVM} achieves the best performance in this scenario (Accuracy = 0.7919, F1 = 0.7575), which shows that training exclusively on benign data supports better generalization to unseen malicious behaviors.
\end{itemize}

Table~\ref{tab:per_class_accuracy_overall} breaks down each model’s accuracy by attack type. Several patterns emerge:

\begin{itemize}

    \item \textbf{Supervised models (MLP \& CNN)} excel on high‐volume attacks and the benign class but fail on rare or subtle threats. Both achieve near‐perfect accuracy on PortScan (MLP: 0.9995; CNN: 0.9713), DDoS (MLP: 0.9993; CNN: 0.9996), FTP‐Patator (0.9987; 0.9987), and BENIGN (MLP: 0.9974; CNN: 0.9822), reflecting strong learning when ample labeled data is available. However, they both struggle on low‐volume Web Attack – Brute Force (MLP: 0.1395; CNN: 0.1860), Web Attack – XSS (MLP: 0.0308; CNN: 0.0231), and Infiltration (MLP: 0.0000; CNN: 0.5714)—indicating overfitting to frequent classes and difficulty generalizing from limited examples. On the three truly novel attacks (DoS slowloris, Slowhttptest, Bot), their accuracy further collapses (\(\le0.37\) on slowloris; \(\le0.08\) on Slowhttptest; 0.00 on Bot), underscoring their reliance on seen patterns.

    \item \textbf{Density‐based LOF} shows notably higher per‐class accuracy on certain novel attack types compared with supervised models. On the Unknown Attack Set, LOF attains accuracy of 0.3176 on DoS slowloris (versus \(\le 0.37\) for MLP/CNN), 0.4150 on Slowhttptest (versus \(\le 0.08\)), and 0.4680 on Bot—demonstrating its ability to flag these outliers more effectively.

    \item \textbf{Boundary‐based OCSVM} stands out on truly novel attacks, achieving moderate accuracy on DoS slowloris (0.57), Slowhttptest (0.94), while maintaining balanced detection across both known and unknown threats (Overall: Accuracy = 0.8356, \(F_1 = 0.5520\); Unknown: Accuracy = 0.7919, \(F_1 = 0.7575\)). This highlights OCSVM’s strength in learning a compact boundary around benign traffic and flagging deviations—even for completely unseen malicious behaviors.

\end{itemize}

\begin{figure}[htbp]
  \centering
  \begin{subfigure}{\linewidth}
    \centering
    \includegraphics[width=\linewidth,height=0.2\textheight,keepaspectratio]{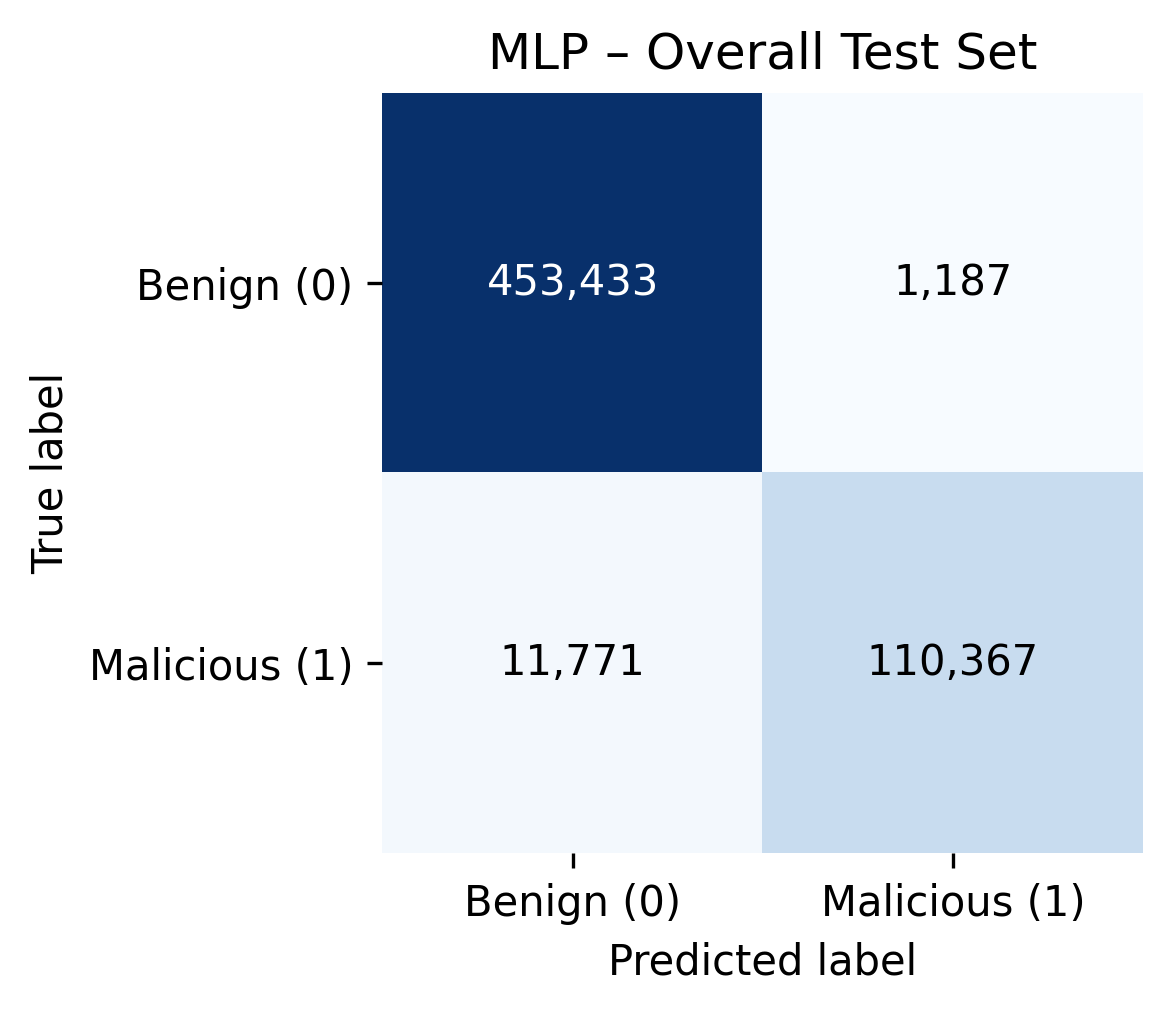}
    \caption{MLP – Overall}
  \end{subfigure}

  \vspace{0.5em}
  \begin{subfigure}{\linewidth}
    \centering
    \includegraphics[width=\linewidth,height=0.2\textheight,keepaspectratio]{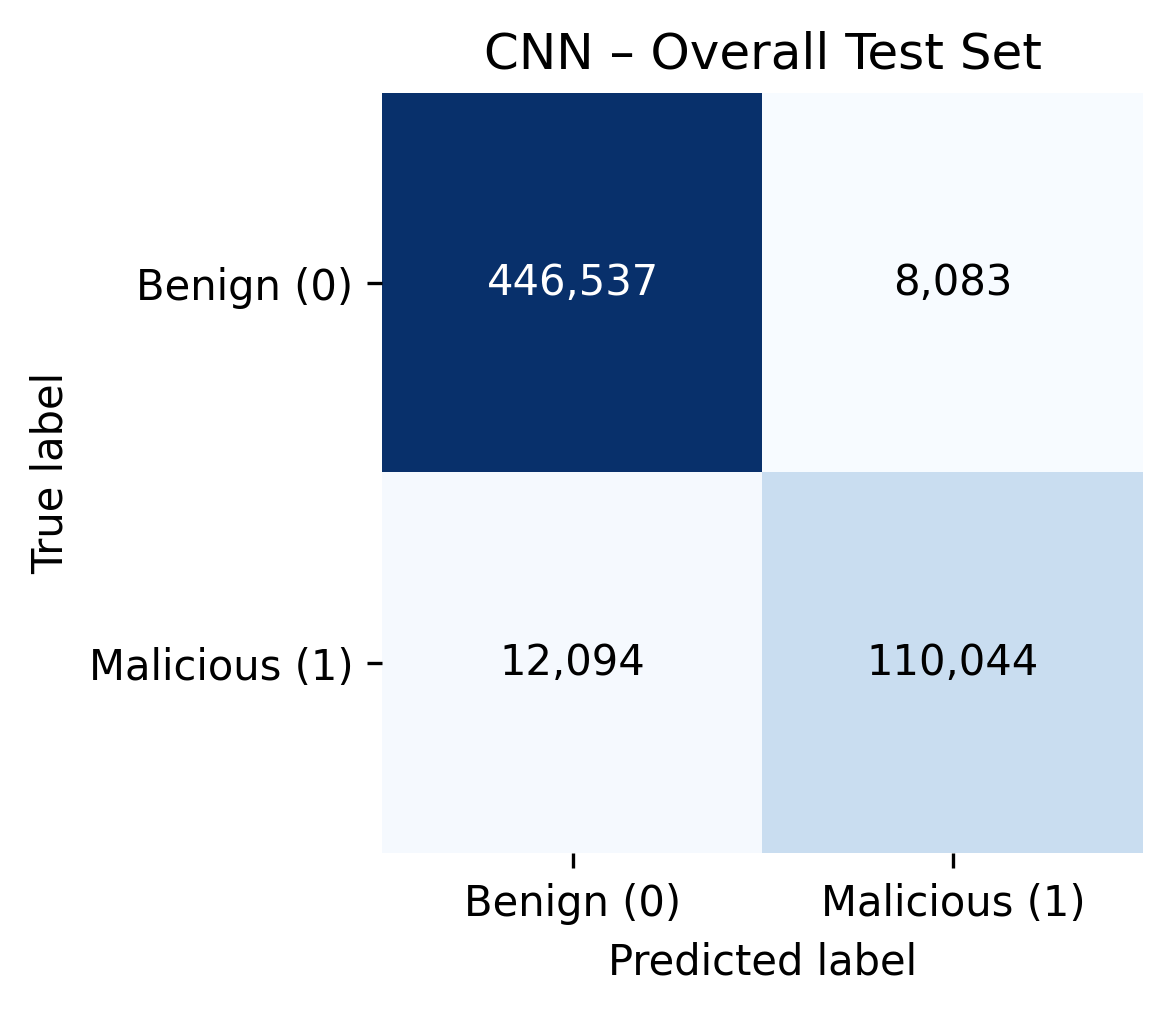}
    \caption{CNN – Overall}
  \end{subfigure}

  \vspace{0.5em}
  \begin{subfigure}{\linewidth}
    \centering
    \includegraphics[width=\linewidth,height=0.2\textheight,keepaspectratio]{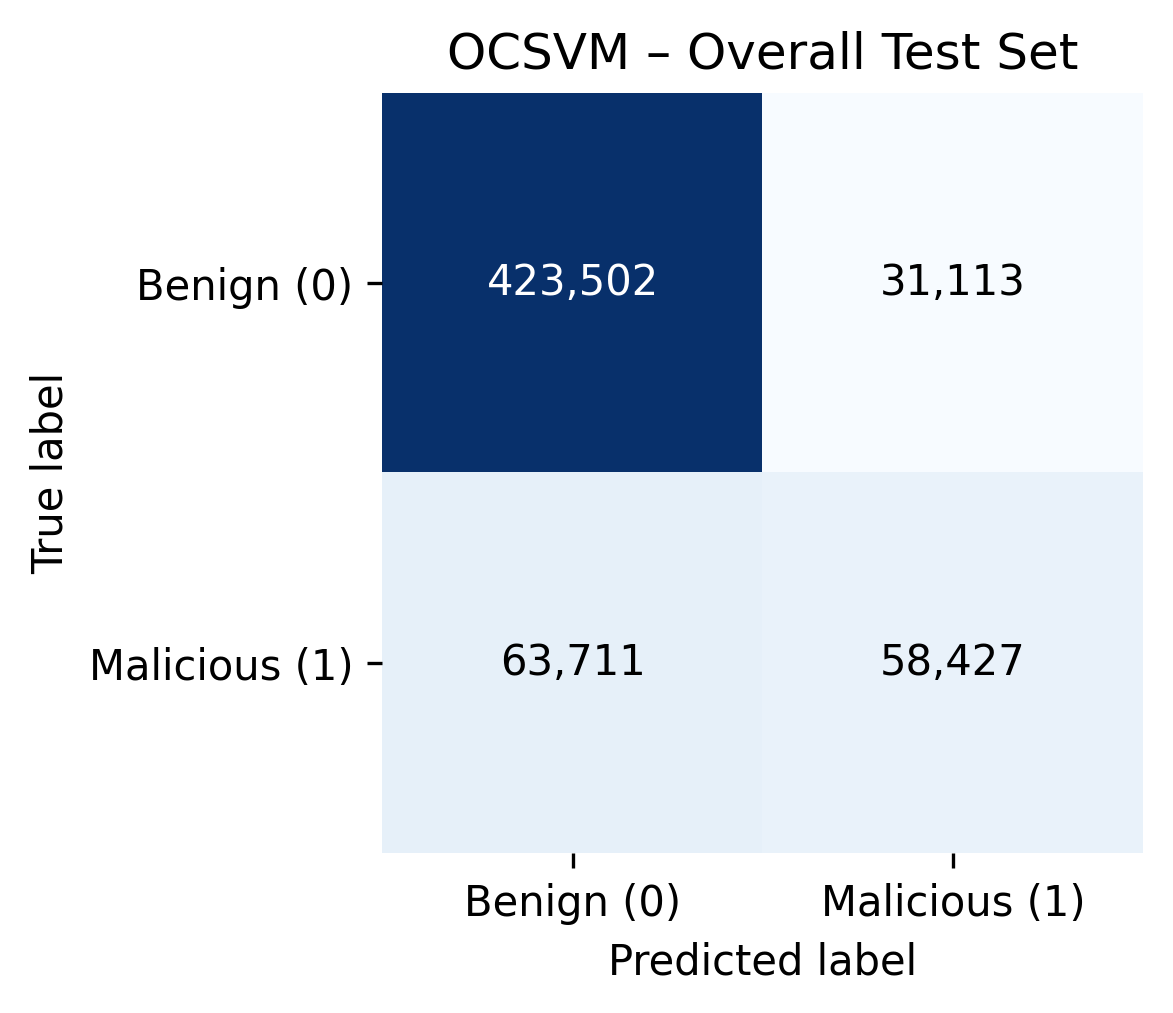}
    \caption{OCSVM – Overall}
  \end{subfigure}

  \vspace{0.5em}
  \begin{subfigure}{\linewidth}
    \centering
    \includegraphics[width=\linewidth,height=0.2\textheight,keepaspectratio]{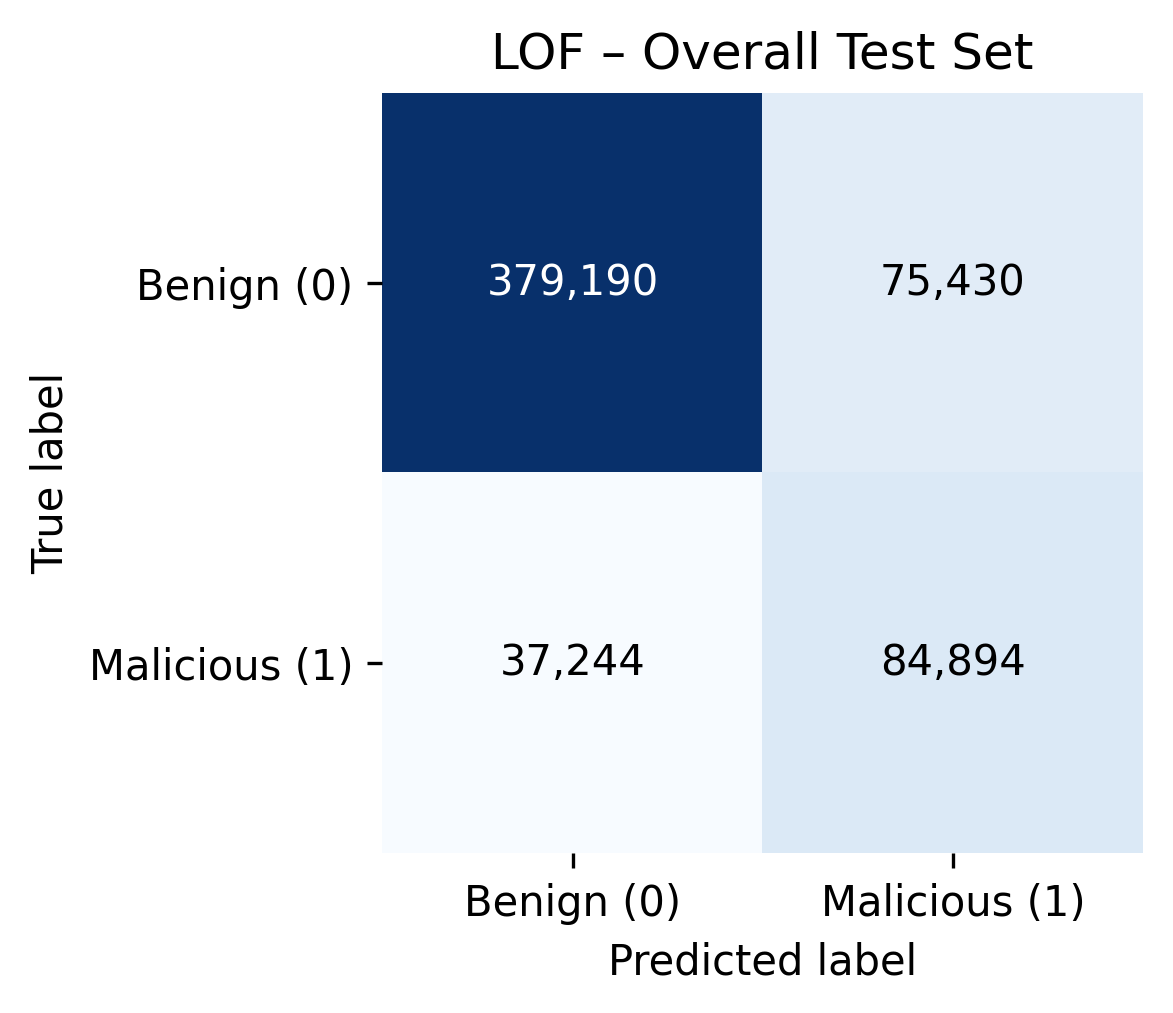}
    \caption{LOF – Overall}
  \end{subfigure}

  \caption{Confusion matrices on the Overall Test Set. (a)–(d) correspond to MLP, CNN, OCSVM, and LOF, respectively.}
  \label{fig:confmat_overall}
\end{figure}
  \label{fig:confmat_overall}

\begin{figure}[htbp]
  \centering
  % Unknown Test Set 四张图，同样设置高度
  \begin{subfigure}[b]{\linewidth}
    \includegraphics[width=\linewidth,height=0.2\textheight,keepaspectratio]{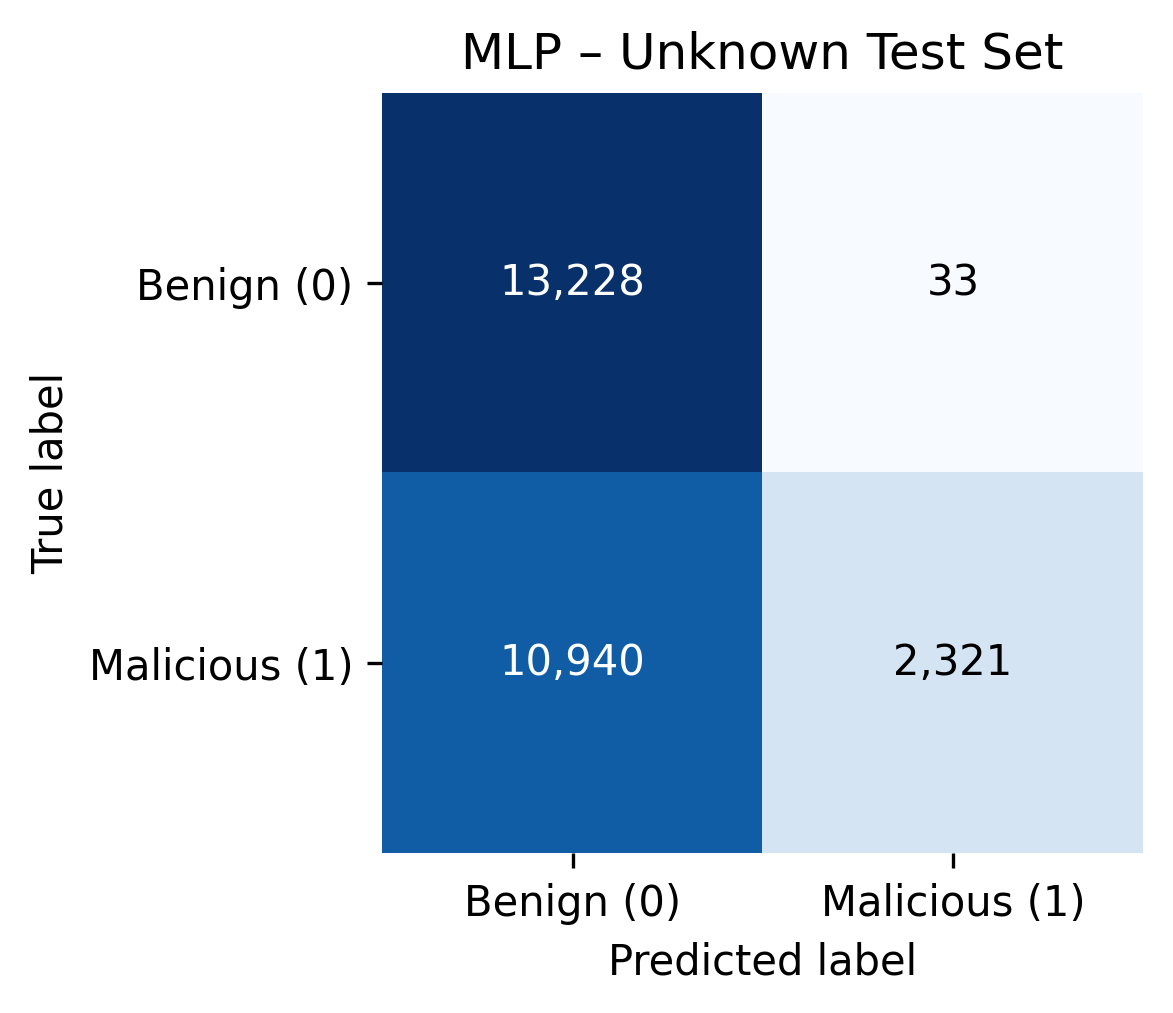}
    \caption{MLP – Unknown}
  \end{subfigure}

  \vspace{0.5em}
  \begin{subfigure}[b]{\linewidth}
    \includegraphics[width=\linewidth,height=0.2\textheight,keepaspectratio]{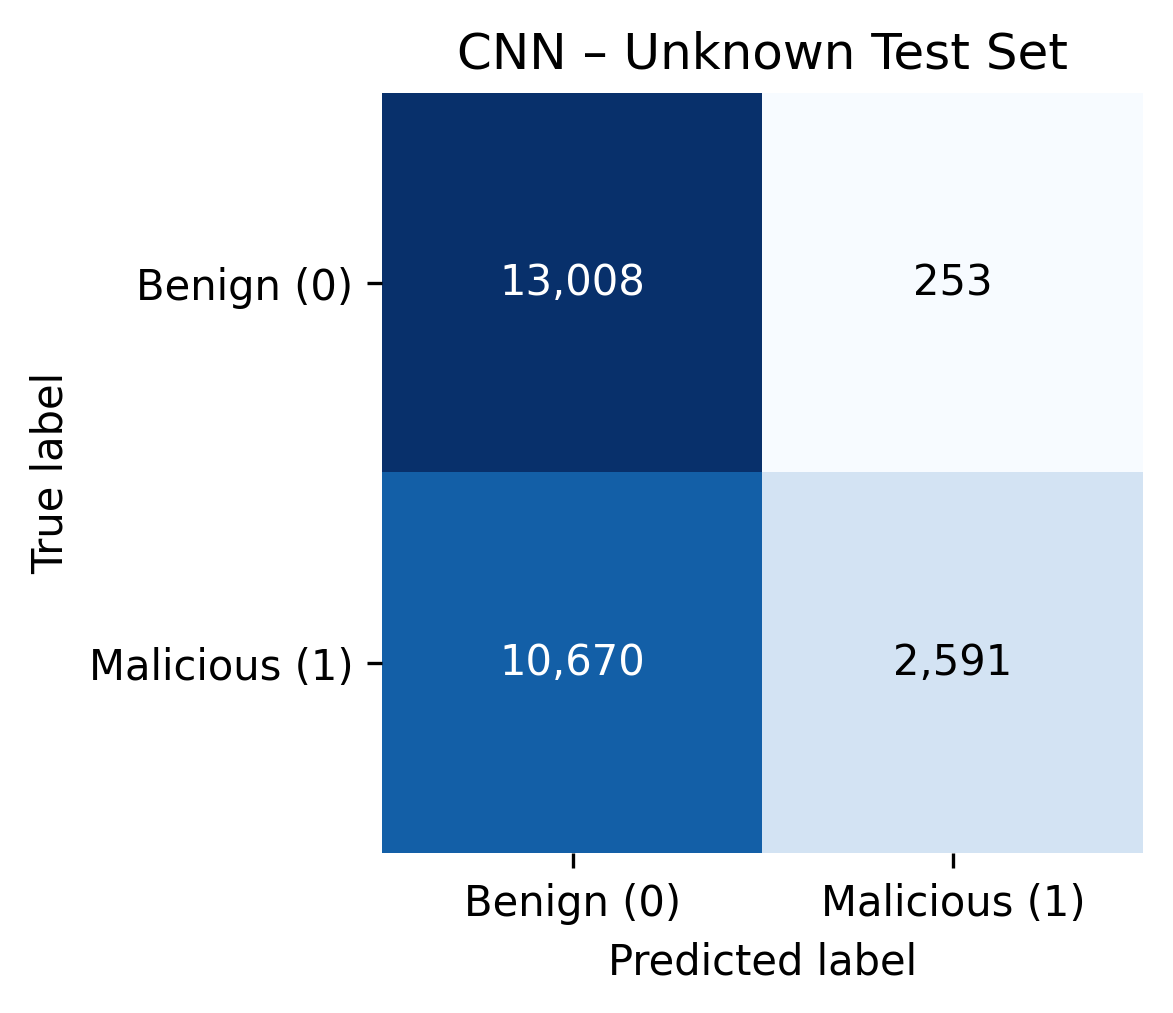}
    \caption{CNN – Unknown}
  \end{subfigure}

  \vspace{0.5em}
  \begin{subfigure}[b]{\linewidth}
    \includegraphics[width=\linewidth,height=0.2\textheight,keepaspectratio]{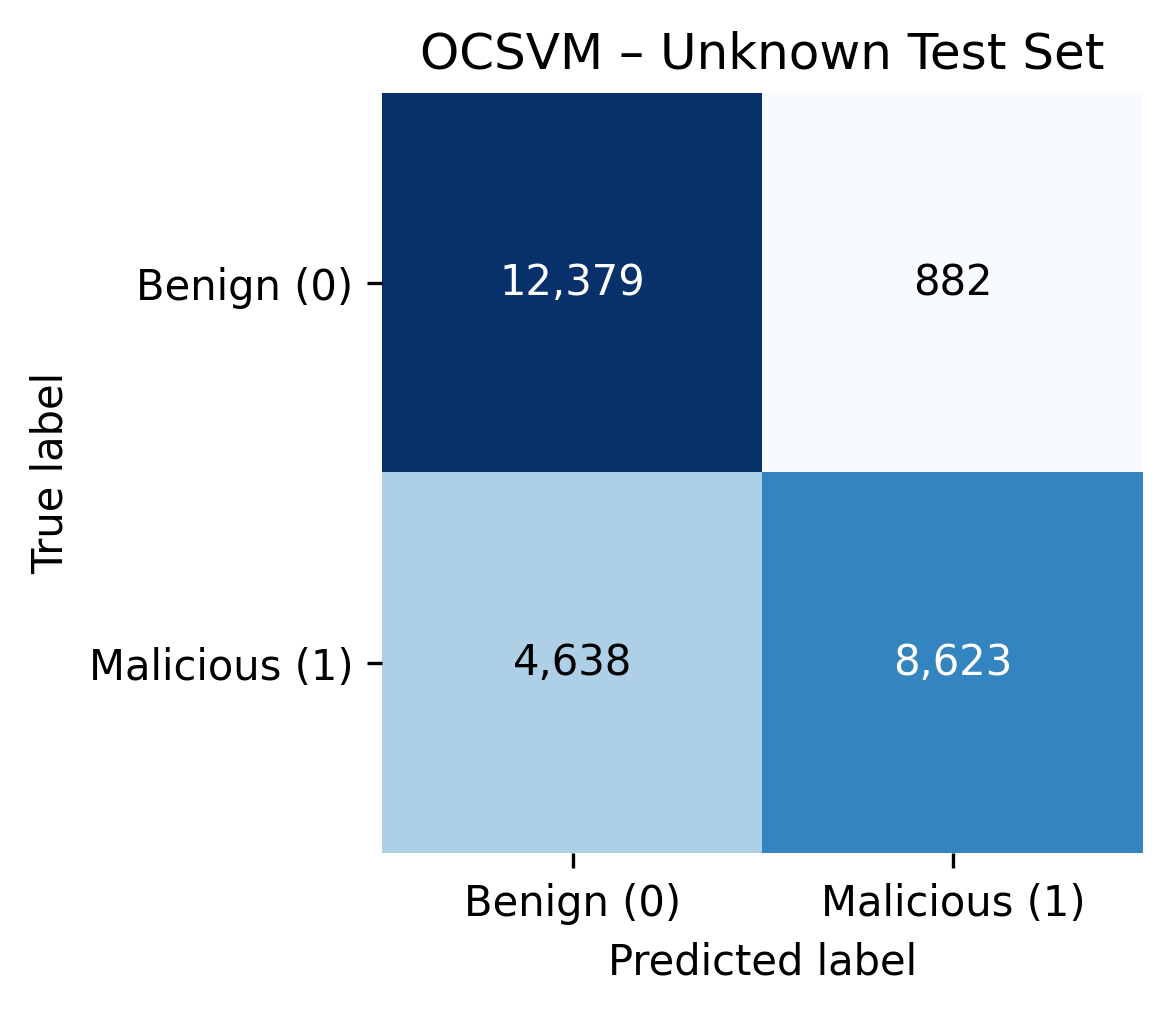}
    \caption{OCSVM – Unknown}
  \end{subfigure}

  \vspace{0.5em}
  \begin{subfigure}[b]{\linewidth}
    \includegraphics[width=\linewidth,height=0.2\textheight,keepaspectratio]{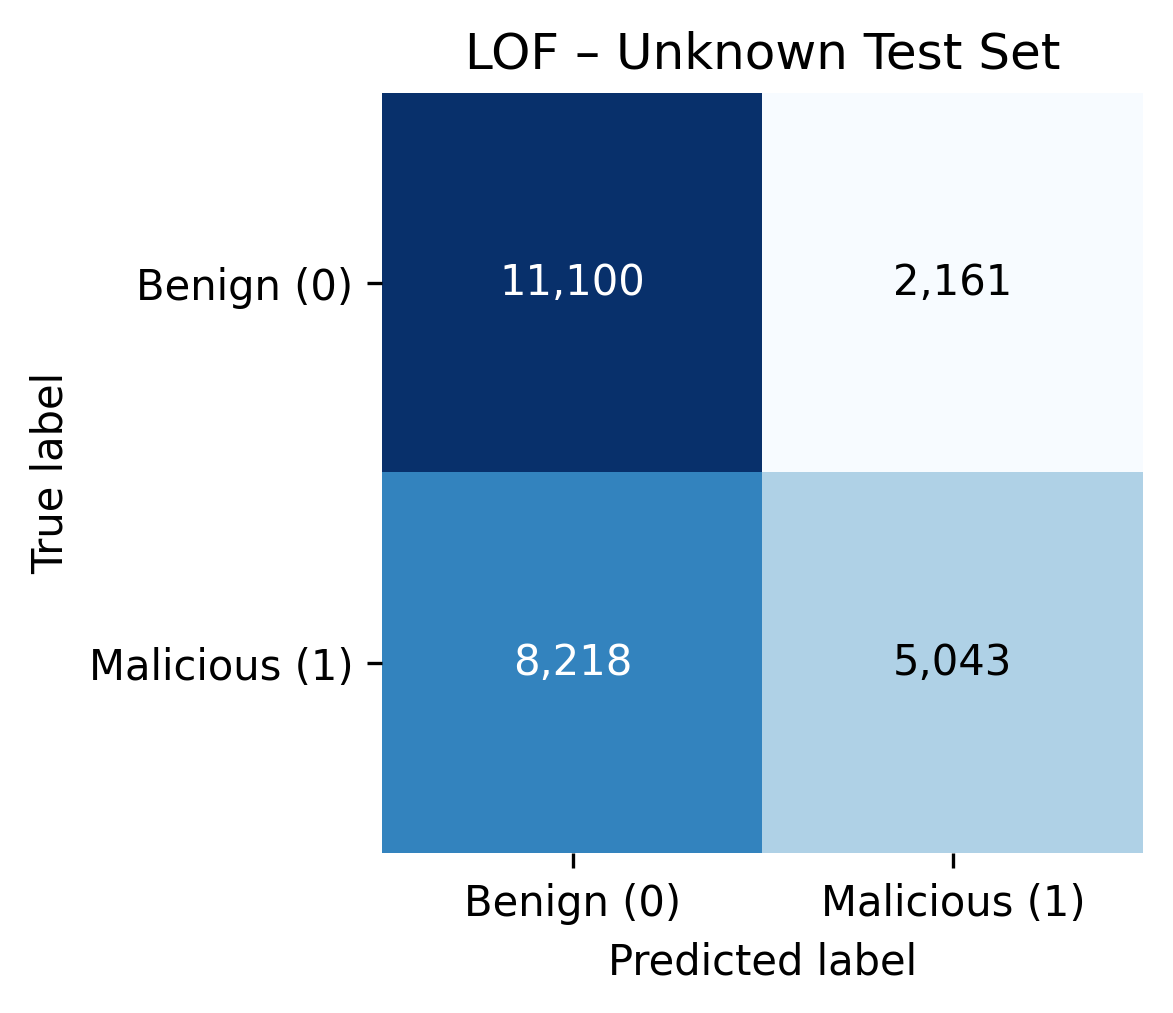}
    \caption{LOF – Unknown}
  \end{subfigure}

  \caption{Confusion matrices on the Unknown Attack Test Set. (a)–(d) correspond to MLP, CNN, OCSVM, and LOF, respectively.}
  \label{fig:confmat_unknown}
\end{figure}

\medskip
\noindent\textbf{Confusion Matrix Analysis.}  
Figure~\ref{fig:confmat_overall} illustrates classifier performance under known attack conditions:

\begin{itemize}
  \item \textbf{MLP} achieves nearly perfect separation, with 453,433 true negatives and 110,367 true positives, suffering only 1,187 false positives and 11,771 false negatives. Such results reflect its high discriminative power when trained on the full range of attack types.
  \item \textbf{CNN} likewise performs strongly, correctly classifying 446,537 benign instances and 110,044 attacks, with 8,083 false positives and 12,094 false negatives—indicating that convolutional feature extraction yields comparable separation to MLP on known threats.
  \item \textbf{LOF} maintains a solid detection ability on known attacks, correctly classifying 379,190 benign instances but incurring 75,430 false positives and missing 37,244 malicious samples—suggesting that density‐based anomaly detection still struggles with perfectly separating benign from malicious traffic in familiar settings.
  \item \textbf{OCSVM} balances these trade-offs more effectively: it significantly reduces false positives to 31,113, albeit with 63,711 false negatives, indicating that unsupervised learning can better delineate benign behaviour with fewer mislabeled benign instances.
\end{itemize}

Figure~\ref{fig:confmat_unknown} shifts focus to the Unknown Attack Test Set:

\begin{itemize}
  \item \textbf{MLP} performance collapses: though it correctly identifies 13,228 benign flows, it mislabels 10,940 malicious samples and only detects 2,321. This stark contrast highlights its inability to generalize beyond the training distribution.
  \item \textbf{CNN} shows a similar collapse: it correctly labels 13,008 benign samples but misclassifies 10,670 attacks, detecting only 2,591, which underscores its limited generalization to truly novel patterns.
  \item \textbf{LOF} shows marked improvement with novel attacks, correctly excluding 11,100 benign flows and flagging 5,043 malicious samples, while incurring 2,161 false positives and missing 8,218 attacks—demonstrating its better adaptability to unseen threats at the cost of moderate misclassification.
  \item \textbf{OCSVM} delivers the most balanced performance in novel scenarios: with only 882 benign samples misclassified and 4,638 attacks missed, it achieves 8,623 true positives. This underscores its superior recall and precision balance in detecting unseen attack types.
\end{itemize}

\noindent\textbf{Summary of Observed Trends:}  
The confusion matrix results underscore a clear pattern: \textbf{MLP and CNN both excel in familiar contexts but severely degrade on new threats}, with CNN showing marginally lower separation power than MLP on known attacks and equally poor generalization to unknown patterns. In contrast, \textbf{LOF offers greater resilience albeit with higher error rates}, and \textbf{OCSVM combines the strengths of both}, maintaining strong detection rates while reducing both false alarms and misses even when facing unknown attacks—demonstrating why it emerges as the most practical approach for real-world intrusion detection.

\section*{\section{Discussion}}

The paradigm-level differences manifest in three key aspects: 

\begin{itemize}
  \item \textbf{MLP} relies on labeled samples of both benign and known attack types during training. Its excellent performance on the Overall Test Set stems from its ability to learn discriminative patterns directly tied to these labels. However, it lacks exposure to unseen attack behaviors and therefore fails to generalize, which explains its drastic drop in recall when tested on novel attacks. Similar generalization challenges have been noted in other systems~\cite{li2023ultrare}, including concept-level backdoor vulnerabilities in interpretable models~\cite{lai2024guarding}.

   \item \textbf{CNN} extends supervised learning with convolutional feature extraction, capturing local temporal and spatial correlations in the data. This leads to performance comparable to MLP on known attacks, but like MLP, it suffers from overfitting to familiar patterns and shows similarly poor generalization to truly novel threats.
  
  \item \textbf{LOF} models normal behavior density without requiring attack labels. It detects anomalies purely based on deviation from established distributions. This grants higher recall on unseen attacks but also induces elevated false positives and false negatives, even for known attacks, reflecting the sensitivity of density thresholds and imperfect boundary estimation.
  
  \item \textbf{OCSVM} trains solely on benign samples, defining a boundary around normal traffic. It flags out-of-distribution patterns as anomalies. This method strikes an optimal balance: it avoids overfitting to known attacks while maintaining precise decision boundaries, resulting in robust detection across both known and novel attack scenarios \cite{kale2023, he2024ddpm, sun2021robust}.
\end{itemize}

\noindent\textbf{Conclusions from Model Behavior:}  
These observations suggest that anomaly detection models defining decision boundaries based only on benign data can be more robust to unseen attacks than supervised models relying on known attack patterns. Among the supervised methods, both \textbf{MLP} and \textbf{CNN} achieve strong performance on familiar threats but suffer significant drops in recall on novel attacks. Within the unsupervised paradigm, our results show that boundary-based \textbf{OCSVM} consistently outperforms density-based \textbf{LOF} for IDS tasks: OCSVM’s global decision surface yields fewer false alarms in mixed-density traffic and more stable recall on novel attacks, making it especially well-suited for real-world intrusion detection, echoing broader efforts to build robust and interpretable AI systems in safety-critical domains such as autonomous driving~\cite{lai2024drive}.

\noindent\textbf{Future Work:}  
Future research could explore hybrid architectures that combine supervised and unsupervised components, leveraging the sequence modeling capability of frameworks like SETransformer \cite{liu2025setransformer, li2025pruning}. Another promising direction involves post-training model adaptation\cite{li2023making}, where attribute unlearning techniques \cite{chen2024post, li2023ultrare, sun2023large}allow intrusion detectors to incorporate new patterns without retraining from scratch.

\vspace{12pt}

\end{document}